\documentclass[aps,prl,superscriptaddress,twocolumn,floatfix,showpacs]{revtex4}
\usepackage{graphics}

\begin{document}

\title{The Collapse of the Spin-Singlet Phase in Quantum Dots}

\author{M. Ciorga}
\affiliation{Institute for Microstructural Sciences, National Research
  Council of Canada, Ottawa, Ontario K1A 0R6, Canada}

\author{A. Wensauer}
\affiliation{Institute for Microstructural Sciences, National Research
  Council of Canada, Ottawa, Ontario K1A 0R6, Canada}
\affiliation{Institute for Theoretical Physics, University of
  Regensburg, D-93040 Regensburg, Germany}

\author{M. Pioro-Ladriere}
\affiliation{Institute for Microstructural Sciences, National Research
  Council of Canada, Ottawa, Ontario K1A 0R6, Canada}
\affiliation{CERPEMA, Universit\'{e} de Sherbrooke, Sherbrooke,
  Qu\'{e}bec J1K 2R1, Canada}

\author{M. Korkusinski}
\affiliation{Institute for Microstructural Sciences, National Research
  Council of Canada, Ottawa, Ontario K1A 0R6, Canada}

\author{Jordan Kyriakidis}
\affiliation{Institute for Microstructural Sciences, National Research
  Council of Canada, Ottawa, Ontario K1A 0R6, Canada}

\author{A.~S. Sachrajda}
\affiliation{Institute for Microstructural Sciences, National Research
  Council of Canada, Ottawa, Ontario K1A 0R6, Canada}

\author{P.~Hawrylak}
\affiliation{Institute for Microstructural Sciences, National Research
  Council of Canada, Ottawa, Ontario K1A 0R6, Canada}

\date{\today}

\begin{abstract}
  We present experimental and theoretical results on a new regime in
  quantum dots in which the filling factor 2 singlet state is replaced
  by new spin polarized phases.  We make use of spin blockade
  spectroscopy to identify the transition to this new regime as a
  function of the number of electrons.  The key experimental
  observation is a reversal of the phase in the systematic oscillation
  of the {\em amplitude} of Coulomb blockade peaks as the number of
  electrons is increased above a critical number.  It is found
  theoretically that correlations are crucial to the existence of the
  new phases.
\end{abstract}


\maketitle

During the last decade, Coulomb blockade (CB) spectroscopic techniques
have been used to investigate the electronic properties of quantum
dots containing a discrete number of
electrons~\cite{1,4,3,2,8,9,10,11}.  CB peaks in the current through
the dot are observed whenever the electrochemical potential of the dot
is aligned with the source and drain leads.  (For lateral dots these
leads are the edges of a two dimensional electron gas, 2DEG.)  The
dot's ground state and the total spin can be tuned by applying a
perpendicular magnetic field.  Each new ground state is observed as a
cusp in the position of the CB peak~\cite{1,4,3,2,8,9,10,11} leading
to the frequent use of this technique for spectroscopy.  The amplitude
of the peaks also contains information.  For irregular quantum dots
with hundreds of electrons, fluctuations in CB peak amplitudes were
used to investigate chaotic phenomena~\cite{chaos} since the amplitude
is reduced whenever the overlap of the dot ground state with the leads
is reduced.  A similar spatial overlap argument was also introduced to
explain drops in the peak amplitude at certain points in the addition
spectrum of medium sized quantum dots~\cite{1}.  For smaller quantum
dots containing fewer electrons, spectroscopic information was,
however, inferred exclusively from the spacing of the CB peaks (the
addition spectrum)~\cite{1,4,3,2,8,9,10,11,12}.  Recently we
discovered an important additional mechanism for amplitude modulation.
In our experiments on lateral devices, we found that electrons
injected into the dot were partially spin polarized~\cite{8,9} (spin
down).  The origin of the observed spin-polarized injection lay in the
exchange-enhanced spin splitting of the magnetic edge states of the
2DEG leads at the entrance and exit barriers of the dot.  The
amplitude modulation is determined by the difference in the electronic
configuration of ground states with two consecutive electron numbers
$N_e$.  Whenever this difference involves a spin up electron the
current is dramatically reduced due to spin blockade~\cite{8,9,10,11}
even if the spatial overlap is large.  Coulomb blockade spectroscopy
of lateral quantum dots is thus accompanied by spin blockade (SB)
spectroscopy~\cite{8,9,10,11}.

We have previously utilized SB spectroscopy to directly investigate
singlet-triplet (ST) transitions~\cite{8,10,11} that occur close to
the filling factor $\nu=2$ regime~\cite{1,4,3} in quantum dots
containing up to 20 electrons.  The $\nu=2$ regime in dots corresponds
to a droplet of electrons occupying an equal number of the lowest
spin-up and spin-down states of the lowest Landau level.  The ST
transitions had been predicted theoretically~\cite{14} and were first
observed in the CB peak spacing of vertical quantum dots by Tarucha
\textit{et al.}~\cite{12} and interpreted in terms of direct and
exchange interactions of the two electrons involved.  In this Letter
we report on a new and unexpected effect which is not discernible in
the spacing of CB peaks but appears clearly in the pattern of CB
amplitude modulation: the complete disappearance or quenching of the
spin-singlet phase itself above a critical number of electrons
$N_{c}$.  We show that this effect can be understood in terms of a
correlated behavior of many electrons.

\begin{figure}
  \resizebox{6.1cm}{5cm}{\includegraphics*{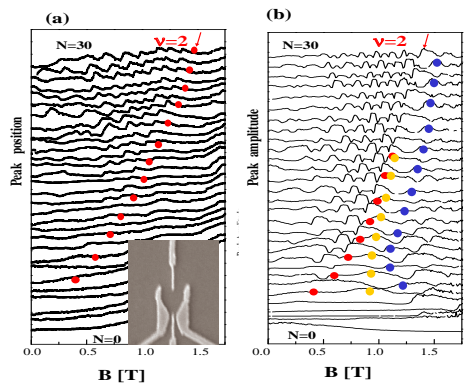}}
  \caption{\label{f1} The addition spectrum (a) and amplitude spectrum
    (b) of the first 30 electrons with the charging energy manually
    removed.  The arrow points to the $\protect\nu=2$ droplet, where
    the spin of the dot oscillates between zero for even electron
    numbers and 1/2 for odd electron numbers.  The inset shows the
    gate layout of the experimental device.}
\end{figure}

The SEM picture of a device similar to the ones used in our
experiments is shown in the inset of Fig.~\ref{f1}a.
The layout of gates in the device allows us to form a slightly
deformed parabolic dot~\cite{15} in which the number of electrons can
be controllably tuned from around 50 electrons down to 1~\cite{11}.
In Fig.~\ref{f1}a we show a typical addition spectrum for the first 30
electrons entering our dot obtained by means of CB spectroscopy.  The
$\nu=2$ line, indicated by a series of red dots and the arrow, is a
very pronounced feature of the spectrum.  Immediately to the right of
this feature is the $\nu=2$ regime.  Figure~\ref{f1}b shows results of
SB spectroscopy, \textit{i.e.}, the {\em amplitude} of the CB peaks
obtained from the same set of measurements as the addition spectrum
shown in Fig.~\ref{f1}a.  The amplitude shows strong oscillations for
$B > 0.4\ \text{T}$ where spin-polarized injection and detection takes
place.  The $\nu=2$ line, marked by red circles, is clearly visible as
a dip in the amplitude starting with five electrons.

On closer inspection, however, it is clear that there are certain
features visible only in the SB spectra.  These are marked with yellow
for even electron numbers and with blue for odd electron numbers.  The
yellow marks approach and eventually cross the line in Fig.~\ref{f1}b,
at a critical number of electrons $N_{c}$, an effect observed in all
of our samples (this feature pictorially marks the transition to the
new phases described in this paper).  The yellow and blue features
correspond to the first spin flip for each $N_e$ as a function of
magnetic field.  It is well established~\cite{1,4,8,10} that as the
field is raised and the filling factor in the quantum dot changes from
$\nu =2$ to $\nu =1$ the quantum dot spin polarizes through a sequence
of spin flips.  It is difficult, however, to experimentally resolve
the first few spin flips using CB spectroscopy.  The amplitude
magneto-fingerprint of a spin flip event (a drop in the peak amplitude
due to both spatial and spin blockades) is found, however, to be
observable even for the first spin flip~\cite{10}.

\begin{figure}
  \resizebox{6.1cm}{!}{\includegraphics*{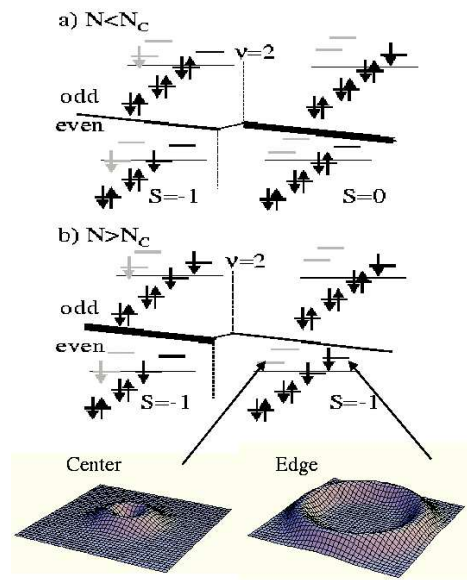}}
  \caption{\label{f2} Electronic configurations of the ground state of
    the $N$-electron droplet in the vicinity of the $\protect\nu=2$
    line for $N < N_c$ (a) and $N > N_c$ (b) (Black: edge orbitals,
    Gray: center orbitals).  Shown schematically is the magnetic field
    evolution of the CB amplitude related to changing the number of
    electrons from even to odd.  Bottom inset shows spatial
    probability density of the center ($m=0$, $n=1$) and edge ($m=9$,
    $n=0$) orbitals. (See text for details.)}
\end{figure}

Firstly we concentrate on the previously understood regime
($N_{e}<25$)~\cite{10}.  The line described by us as the $\nu =2$ line
is more accurately defined as the low-field boundary of the $\nu =2$
singlet phase.  To the right of the $\nu =2 $ line electrons occupy
spin-split states $(m,0)$ of the lowest Landau level with positive
angular momentum $m$~\cite{8} (the zero indicates the lowest Landau
level).  The states $(m,0)$ are equally spaced, with an energy spacing
decreasing with increasing magnetic field which would eventually
converge at very high magnetic fields to a single Landau level.  For
each state $(m,0)$ there is a state $(m,1)$ originating from the
second Landau level separated from it by a fixed energy which would
approach the cyclotron energy at very high magnetic fields.  For an
even number of electrons $N_{e}=2N$ both spin down and up states
$(m,0)$ are filled up to the Fermi level forming the $\nu =2$
spin-singlet (total spin $S=0$) state.  This state, together with
electronic configurations of other states in close proximity to the
$\nu =2$ line, are shown schematically in Fig.~\ref{f2}a.
For an odd number of electrons, one unpaired electron occupies a level
at the edge of the droplet and the total spin of the in this case is
$-1/2$.  As the magnetic field is lowered, the singlet phase becomes
unstable against the transfer of an electron from the edge of the dot
to the second Landau level orbital (0,1) in the center of the dot.
The spatial charge distribution corresponding to the center and a
representative edge orbital are shown at the bottom of Fig.~\ref{f2}.
For an even number of electrons, decreasing the magnetic field
transfers an electron from an edge to a center orbital with angular
momentum $-1$ while simultaneously flipping its spin.  The dot is then
in a triplet state formed by one electron in the edge and one in the
center of the droplet.  This configuration does not, of course, just
consist of two electrons, but is a many-body state~\cite{10,15,14}.
For an odd number of electrons, an unpaired electron at the edge of
the droplet is also transferred to the center but without flipping its
spin so the total spin of the droplet in this case remains
unchanged~\cite{10,15,14}.  As seen in Fig.~\ref{f1}a, there is no
discernible experimental difference in the magnetic field at which
transitions for even and odd total electron numbers take place.  At
higher magnetic fields there is a second boundary.  The $\nu =2$
droplet becomes unstable against spin flips at the edge of the quantum
dot.  This boundary is dependent, however, on whether the dot contains
an even or odd number of electrons.  For both odd and even electron
numbers $N_e$, a spin up electron at the edge of the droplet moves to
the first available empty orbital with a higher angular momentum and
flips its spin.  For even $N_{e}$, and in the absence of interactions,
the electron flips its spin at the edge whenever the cost of kinetic
energy is compensated by the gain in Zeeman energy $E_{z}$.  For odd
$N_{e}$, spin flips cost twice as much kinetic energy.  Hence, the
magnetic field for spin flips for odd $N_{e}$ is much higher than that
for even $N_{e}$ ---the parity, not the magnitude, of $N_e$ is what is
important.  These features can be seen in the data corresponding to
the first electron spin flip.  The weak dependence on $N_e$ together
with the large shift between the even and odd spin flips, while
renormalized by interactions, is still visible in the SB spectrum of
quantum dot shown in Fig.~\ref{f1}.

\begin{figure}
  \resizebox{6.1cm}{!}{\includegraphics*{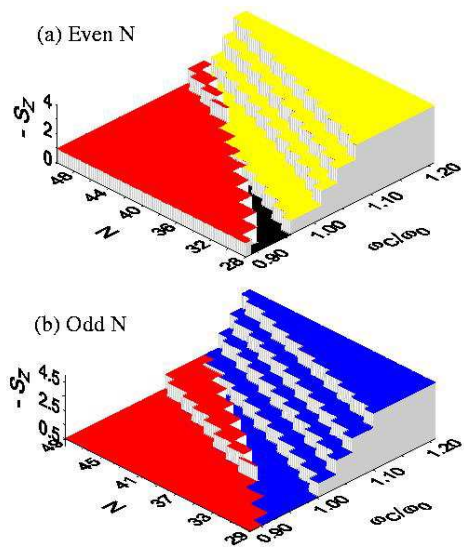}}
  \caption{\label{f3} Calculated ground-state spin configurations of
    electron droplets with even- and odd-$N$ near the collapse of the
    spin-singlet phase.}
\end{figure}

A number of model calculations for our system---spin density
functional theory in the local spin-density approximation (LSDA) with
Landau-level mixing, and Hartree-Fock calculations with and without
Landau-level mixing---were performed.  In particular, the stability of
the $\nu=2$ singlet phase against transfer of electrons to the center
(center configurations) and against spin flips at the edge (edge
configurations) was studied.  In Fig.~\ref{f3} we show the calculated
spin of the ground state configuration as a function of magnetic field
for electron droplets with even and odd $N_e$.
Red denotes center configurations, yellow denotes edge configurations
for even $N_e$, blue denotes edge configurations for odd $N_e$, and
black denotes the $\nu=2$ spin singlet droplet.  The self-consistent
calculations employed the LSDA and include mixing of ten Landau
levels.  The calculations used a confinement energy of $\omega=1$~meV
(extracted from the magnetic-field evolution of the CB peak
corresponding to the first electron in the dot), a Zeeman energy of
$E_z=0.04$~meV/T, and strictly 2D Coulomb interactions.  The results
of these calculations were already schematically summarized in
Fig.~\ref{f2}.  The LSDA and Hartree Fock calculations with and
without a mixing of Landau levels all give a finite stability range of
the spin singlet droplet (black region).  At a critical number of
electrons $N_c$, the spin-singlet $\nu=2$ phase ceases to be the
ground state.  As seen in Fig.~\ref{f3}, upon increasing the field,
the dot with even $N_e$ evolves from a center configuration with a
spin-down electron at the center and at the edge, to an edge
configuration with two spin-down electrons at the edge of the droplet.
This is shown schematically in Fig.~\ref{f2}b.  A comparison of
Figs.~\ref{f2}a, \ref{f2}b, and Fig.~\ref{f3} reveals a change in the
center configuration of the droplet consisting of an odd number of
electrons.  The key effect is the triggering of spin polarization at
the edge by the spin and charge of an electron at the center.  These
configurations persist over a finite range of electron numbers, as
shown in Fig.~\ref{f3}.

\begin{figure}
  \resizebox{6.1cm}{!}{\includegraphics*{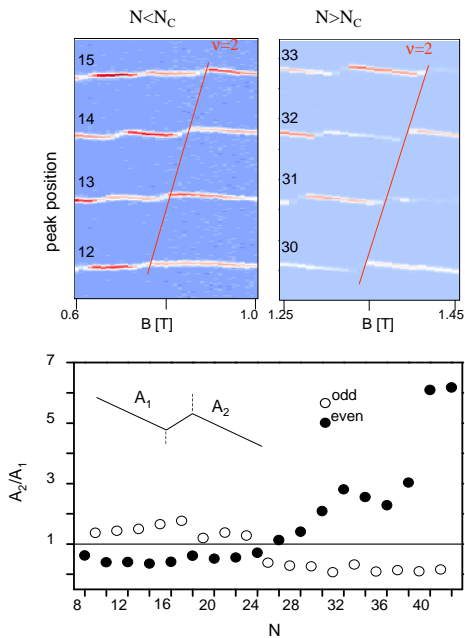}}
  \caption{\label{f4} Upper panel: inverted gray-scale showing
    magnetic field evolution of four CB peaks in the vicinity of the
    $\protect\nu=2$ line for $N<N_c$ and $N>N_c$.  Lower panel: ratio
    of CB peak amplitude $A_2$ on the right of $\protect\nu=2$ line to
    CB peak amplitude $A_1$ on the left as a function of electron
    number.}
\end{figure}

This effect is too weak to be observed in the addition spectrum of
Fig.~\ref{f1}.  In contrast, the effect \emph{can} be directly
observed with SB spectroscopy.  Consider how the predicted changes in
the ground-state configurations should affect the current through the
dot.  The magnetic-field evolution of the CB peaks is schematically
shown in Fig.~\ref{f2}.  The thickness of the lines indicates the
expected current amplitude (thin for low, thick for high).  For $N_e <
N_c$, the two center configurations for odd and even $N_e$ differ by
one spin up electron at the edge of the dot.  The two edge
configurations differ by a spin down electron at the edge.  Because of
spin-polarized injection~\cite{11}, SB-spectroscopy measurements
should reveal a relatively small current flowing through the dot
whenever a transition occurs between center configurations (left of
the $\nu=2$ line) and a relatively large current whenever a transition
occurs between edge configurations (right of the $\nu=2 $ line).  A
similar analysis for transitions from odd to even $N_e$ would give
high current on the left of $\nu=2$ line and low current on the
right~\cite{11}.  In the case of $N_e > N_c $, as seen in
Fig.~\ref{f2}b, the electronic configurations of the respective ground
states have changed.  The initial center configuration for even $N_e$
and the final center configuration for odd $N_e$ differ by a spin down
electron at the edge, and so we expect a large current on the left of
the $\nu=2$ line.  The initial and final edge configurations differ by
a spin-up electron at the edge of the droplet, and so the observed
current is expected to be low.  Thus, the collapse of the $\nu=2$
spin-singlet droplet should be seen through SB spectroscopy as a
reversal of the amplitude oscillation pattern in the vicinity of the
$\nu=2$ line as the number of electrons is increased.  This is indeed
observed in our experiments.  In Fig.~\ref{f4}, we show inverted
gray-scales of the magnetic-field evolution of four CB peaks in the
vicinity of the $\nu=2$ line in the regime of both $N_e<N_c$ and
$N_e>N_c$.
Dark and light shades in the gray-scale indicate respectively large
and small peak amplitude.  The amplitude of the CB peaks behaves in
the way predicted in the above discussion of ground-state electronic
configurations.  In the bottom panel of Fig.~\ref{f4}, we plot the
ratio of the peak amplitude $A_2$ on the right side of the $\nu=2$
line to the amplitude $A_1$ on the left side of the $\nu=2$ line as a
function of electron number $N_e$.  For a low electron number, this
ratio is greater than unity when adding an odd electron to the dot and
less than unity when adding an even electron.  The pattern reverses
around $N_c=25$.  This number is different from the calculated one
which perhaps points to our overestimation of the strength of Coulomb
interactions, and the lack of detailed knowledge of the change of
confinement on the number of electrons.  The collapse of the $\nu=2$
spin-singlet droplet and the critical number of electrons $N_c$
observed in experiment was reproduced by Hartree-Fock calculations for
a number of confining energies.  However, only LSDA calculations which
include correlations were capable of producing a phase diagram leading
to amplitude reversal.  Hence amplitude reversal appears to be
connected to correlations, and more realistic calculations are in
progress to illuminate this connection.

To summarize, we have studied the stability of the $\nu=2$
spin-singlet phase of a quantum dot as a function of electron number
$N_e$ and magnetic field $B$.  We have demonstrated that this phase
collapses at a certain electron number $N_c$ in favor of
spin-polarized configurations.  We were able to observe this effect
experimentally with spin-blockade spectroscopy.  The experiments and
calculations demonstrate new effects uncovered by the control of
electron spin in a nanoscale object with a tunable and controlled
number of electrons.  These findings should have impact on the merging
fields of spintronics, nanotechnology, and quantum information, which
require the ability to control and manipulate spin and charge at the
single-electron level.

\end{document}